\documentclass[twocolumn,prl,aps,showpacs,floatfix,superscriptaddress,preprintnumbers,amsmath,amssymb]{revtex4}

\usepackage{graphicx}
\usepackage{amsmath}
\usepackage{color}

\begin{document}
\newcommand{\mm}{MiniMuPAD}
\newcommand{\mfs}{Mn$_{1-x}$Fe$_{x}$Si}
\newcommand{\mcs}{Mn$_{1-x}$Co$_{x}$Si}
\newcommand{\fcs}{Fe$_{1-x}$Co$_{x}$Si}

\newcommand{\rxx}{$\rho_{xx}$}
\newcommand{\rxy}{$\rho_{xy}$}
\newcommand{\rxytop}{$\rho_{\rm xy}^{\rm top}$}
\newcommand{\Drxyt}{$\Delta\rho_{\rm xy}^{\rm top}$}
\newcommand{\Sxy}{$\sigma_{xy}$}
\newcommand{\Sxya}{$\sigma_{xy}^A$}

\newcommand{\bco}{$B_{\rm c1}$}
\newcommand{\bct}{$B_{\rm c2}$}
\newcommand{\bao}{$B_{\rm A1}$}
\newcommand{\bat}{$B_{\rm A2}$}
\newcommand{\beff}{$B^{\rm eff}$}

\newcommand{\btr}{$B^{\rm tr}$}

\newcommand{\tc}{$T_{\rm c}$}
\newcommand{\ttr}{$T_{\rm tr}$}

\newcommand{\mb}{$\mu_0\,M/B$}
\newcommand{\dmdb}{$\mu_0\,\mathrm{d}M/\mathrm{d}B$}
\newcommand{\ddmddb}{$\mathrm{\mu_0\Delta}M/\mathrm{\Delta}B$}
\newcommand{\cm}{$\chi_{\rm M}$}
\newcommand{\cac}{$\chi_{\rm ac}$}
\newcommand{\rechi}{${\rm Re}\,\chi_{\rm ac}$}
\newcommand{\imchi}{${\rm Im}\,\chi_{\rm ac}$}

\newcommand{\ozz}{$\langle100\rangle$}
\newcommand{\ooz}{$\langle110\rangle$}
\newcommand{\ooo}{$\langle111\rangle$}
\newcommand{\too}{$\langle211\rangle$}

\newcommand{\cp}{CryoPAD}
\newcommand{\mup}{MuPAD}


\title{Critical spin-flip scattering at the helimagnetic transition of MnSi}

\author{J. Kindervater}
\affiliation{Physik-Department E21, Technische Universit\"at M\"unchen, D-85748 Garching, Germany}

\author{W. H\"au\ss ler}
\affiliation{Physik-Department E21, Technische Universit\"at M\"unchen, D-85748 Garching, Germany}
\affiliation{Heinz Maier-Leibnitz Zentrum, Technische Universit\"at M\"unchen, D-85748 Garching, Germany}

\author{M. Janoschek}
\affiliation{Los Alamos National Laboratory, Los Alamos, New Mexico 87545, USA}

\author{C. Pfleiderer}
\affiliation{Physik-Department E21, Technische Universit\"at M\"unchen, D-85748 Garching, Germany}

\author{P. B\"oni}
\affiliation{Physik-Department E21, Technische Universit\"at M\"unchen, D-85748 Garching, Germany}

\author{M. Garst}
\affiliation{Institute for Theoretical Physics, Universi\"at zu K\"oln, D-50937 K\"oln, Germany}

\date{\today}

\begin{abstract}
We report spherical neutron polarimetry (SNP) and discuss the spin-flip scattering cross sections as well as the chiral fraction $\eta$ close to the helimagnetic transition in MnSi. For our study, we have developed a miniaturised SNP device that allows fast data collection when used in small angle scattering geometry with an area detector. Critical spin-flip scattering is found to be governed by chiral paramagnons that soften on a sphere in momentum space. Carefully accounting for the incoherent spin-flip background, we find that the resulting chiral fraction $\eta$ decreases gradually above the helimagnetic transition reflecting a strongly renormalised chiral correlation length with a temperature dependence in excellent quantitative agreement with the Brazovskii theory for a fluctuation-induced first order transition.
\end{abstract}


\vskip2pc
\maketitle

It has long been established that B20 transition metal compounds support a well-understood hierarchy of energy scales \cite{Landau} comprising in decreasing strength ferromagnetic exchange, Dzyaloshinsky-Moriya spin-orbit interactions and, finally, magnetic anisotropies due to higher-order spin-orbit coupling. In turn, zero temperature spontaneous magnetic order in these materials appears essentially ferromagnetic on short distances with a long wave-length homochiral twist on intermediate distances that propagates along directions favoured by the magnetic anisotropies on the largest length scales.

In recent years the nature of the associated helimagnetic transition at $T_{\rm c}$, which makes contact with areas ranging from nuclear matter over quantum Hall physics to soft matter systems, has been the topic of heated scientific controversy since critical helimagnetic spin fluctuations are able to drive the transition first order. However, considerable differences exists as to the proposed character of these fluctuations. Based on a minimal description taking only into account the three scales mentioned above two scenarios may be distinguished. First, according to Bak and Jensen \cite{Bak1980} when the magnetic anisotropies are sufficiently strong, anisotropic critical paramagnons develop along the easy axis already at $T>T_{\rm c}$. Second, in the opposite limit when the magnetic anisotropies are weak, critical spin fluctuations soften isotropically on the surface of a sphere in momentum space giving rise to a fluctuation-driven first order transition in the spirit of a proposal by Brazovskii \cite{Brazovskii1975}. In addition, a third and completely different scenario by R{\"o}{\ss}ler, Bogdanov and Pfleiderer \cite{Rossler2006} has generated great interest, which requires, however, an additional phenomenological parameter beyond the minimal model. In this model the generic formation of a skyrmion liquid phase between the paramagnetic and helimagnetic state is predicted, which implies an additional phase transition at a temperature $T_{\rm sk}>T_{\rm c}$. 

Eearly experimental studies of the electrical resistivity \cite{Petrova:PRB2006}, specific heat \cite{Stishov:PRB2007}, thermal expansion \cite{Stishov:JPCM2008,Petrova:PRB2010}, ultrasound attenuation \cite{Stishov:JPCM2009} and neutron scattering \cite{Grigoriev:PRB2010} in MnSi, as the most extensively studied B20 compound, were interpreted in terms of the scenario by Bak and Jensen. Comprehensive elastic neutron scattering, demonstrating  critical fluctuations on the surface of a sphere, together with specific heat, susceptibility and magnetisation measurements, recently changed this view, providing quantitatively consistent evidence of a fluctuation-driven first order transition as proposed by Brazovskii \cite{janoschek2013fluctuation,Bauer:PRB2012,Bauer:PRL2013,Buhrandt:PRB2013}, see also Ref.~\cite{Zivkovic:arxiv2014} for related work on Cu$_2$OSeO$_3$.

Following the discovery of a skyrmion lattice phase in small applied magnetic fields just below $T_c$ \cite{Muehlbauer2009,Muenzer:PRB2010,Yu:Nature2010,Yu:NatureMaterials2011}, several authors have argued that the specific heat and susceptibility provide evidence for further complex phases including a skyrmion liquid phase at zero field \cite{Hamann2011,Samatham:PSSB2013,Wilhelm:PRL2011,Wilhelm:JPCM2012,Cevey:PSSB2013}. Most importantly, it has been claimed that the observation of a chiral fraction $\eta \approx 1$ up to at least 1\,K above $T_c$ in a seminal SNP study in MnSi by Pappas et al. \cite{Pappas2009,Pappas2011} provides microscopic evidence supporting a skyrmion liquid phase. However, as explained in our Letter $\eta$ provides a measure of the asymmetry of magnetic spin-flip scattering, assuming extreme values $\eta = \pm 1$ if one of the spin-flip scattering processes, i.e., $\uparrow$ to $\downarrow$ or vice versa, is forbidden. Hence $\eta$ shows to what extent a magnetic system is homochiral. In contrast, by definition $\eta$ is neither a direct measure of the topological winding as the defining new property of the skyrmion liquid, nor of the phase relationship of the underlying multi-$q$ modulations determined recently in the skyrmion lattice phase in MnSi \cite{Adams2011}. Moreover, to the best of our knowledge a theoretical link between $\eta$ and the formation of a skyrmion liquid phase has not been reported either. 

Motivated by the broad interest in the helimagnetic transition of chiral magnets and the special attention paid to $\eta$ we have revisited the entire issue from a more general point of view in an experimental and theoretical study of the critical spin-flip scattering in MnSi. We thereby do not find any evidence suggesting an additional phase transition above $T_{\rm c}$. As our main conclusion, our SNP results are in excellent \textit{quantitative} agreement with the minimal model of a fluctuation-induced first order transition as predicted by Brazovskii, establishing also quantitative consistency with previous specific heat, magnetisation, susceptibility and elastic neutron scattering studies \cite{janoschek2013fluctuation,Bauer:PRB2012,Bauer:PRL2013,Buhrandt:PRB2013}.

Carefully considering the experimental requirements to go beyond previous SNP studies, presented in detail below, revealed as most prominent aspect the need to track incoherent signal contributions. To meet these requirements we have developed a versatile miniaturised SNP device \cite{Jonas-diploma}, which in its present version offers great flexibility at scattering angles up to $15^{\circ}$. In particular, as opposed to the large size of SNP devices such as {\cp} or {\mup} \cite{tasset1989zero,janoschek2007spherical} our entire set-up (diameter 50\,mm; height 120\,mm) is integrated into a normal sample stick fitting a standard pulse-tube cooler. In turn this reduces the time required for setting up our SNP device to the time needed for a conventional sample change. Moreover, when combined with an area detector fast data collection at various sample orientations and momentum transfers are readily possible.


\begin{figure}
 \includegraphics[width=0.45\textwidth]{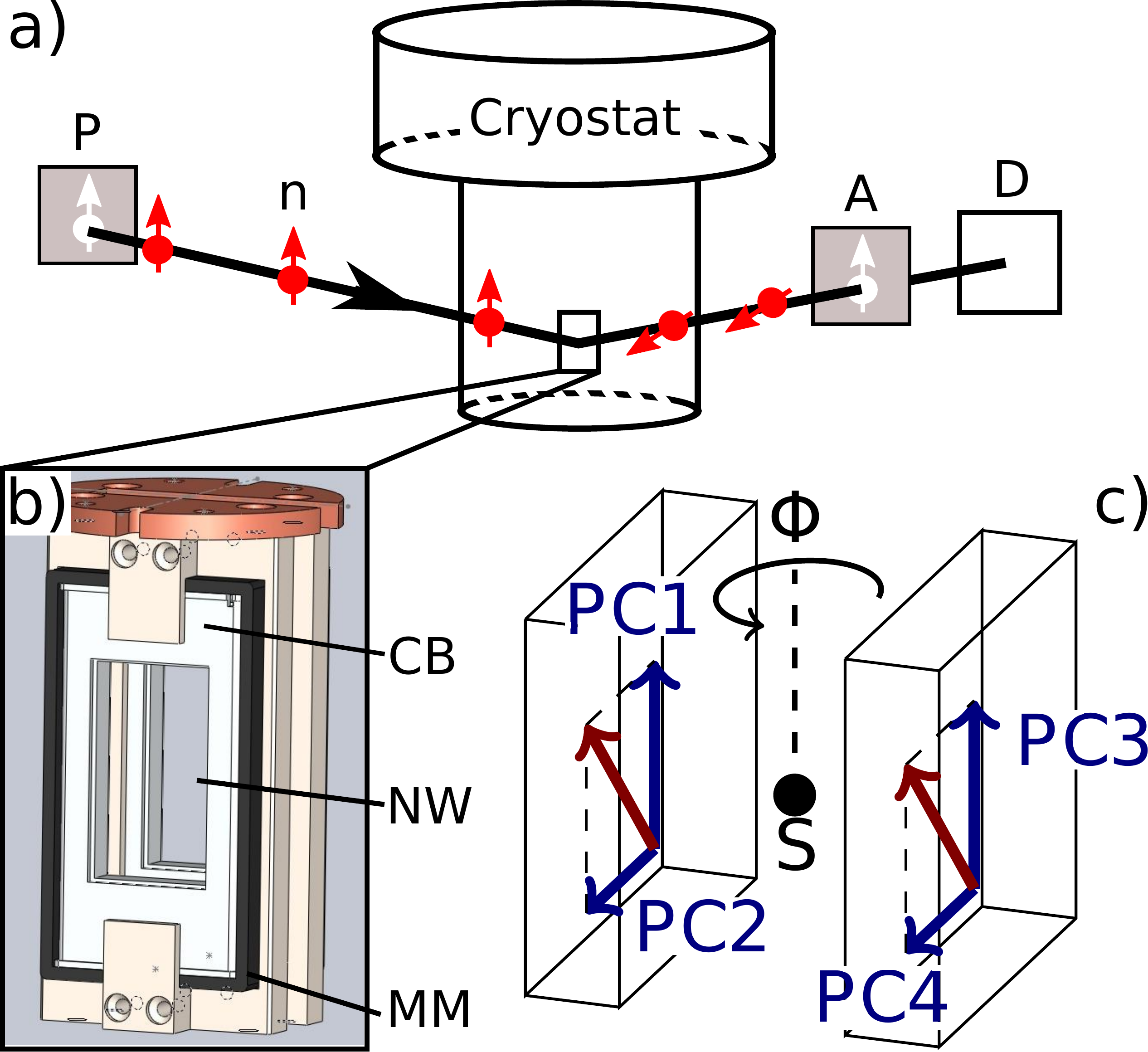}
 \caption{Schematic depiction of the miniaturised SNP device. (a) Schematic overview of the complete setup with cryostat, polarizer (P), analyzer (A) and detector (D). (b) Close-up view of the SNP device, as composed of the coil bodies (CB) with their neutron window (NW). The coils are surrounded by a mu-metal yoke (MM). (c) Orientation of the precession coils (blue arrows), local magnetic field (red arrow) and sample (S).}
 \label{figure1}
\end{figure}

Shown in Fig.\,\ref{figure1}(a) is a schematic depiction of the miniaturised SNP set-up developed for our study. Pairs of crossed precession coils (PC) before and after the sample generate Larmor precessions of the neutron polarisation that permit to rotate the polarisation of an incoming and scattered neutron beam in any arbitrary direction. The precession coils are wound on coil bodies (CB) with a large neutron window (NW), see Fig.\,\ref{figure1}(b). As its main advantage parasitic rotations of the polarisation are minimised by the very compact geometry and the mu-metal yokes (MM) around the precession coils, which short-circuit both external fields and the precession fields.

The measurements reported in this Letter were performed on the beam-line RESEDA \cite{haussler2007reseda} at the Heinz Maier-Leibnitz Zentrum of the Technische Universit\"at M\"unchen. Neutrons were polarised with a cavity providing a polarisation of 95\,\% and analysed with a bender at an efficiency of 98\,\%. Data were recorded at a neutron wavelength $\lambda=4.5\,{\rm \AA}$ with a wavelength spread $\Delta\lambda/\lambda=0.16$. Using a CASCADE \cite{haussler2011detection} area detector (PSD) with $200\,{\rm mm}\times200\,{\rm mm}$ active area ($128 \times 128$ pixels) at a distance to the sample of 1596\,mm our set-up corresponded effectively to small angle neutron scattering. The FWHM of the resolution and the wavelength band $\sigma_{\textrm{fCol}}=3.6\cdot10^{-3}\,\textrm{\AA}^{-1}$ and $\sigma_{\textrm{W}}=2.6\cdot10^{-3}\,\textrm{\AA}^{-1}$, respectively, as considered at the wave vector of the helical modulation,  $k=0.039$\,\AA\, of MnSi, resulted in a momentum uncertainty $\Delta Q/Q\approx10\,\%$.

\begin{figure}
\includegraphics[width=0.45\textwidth]{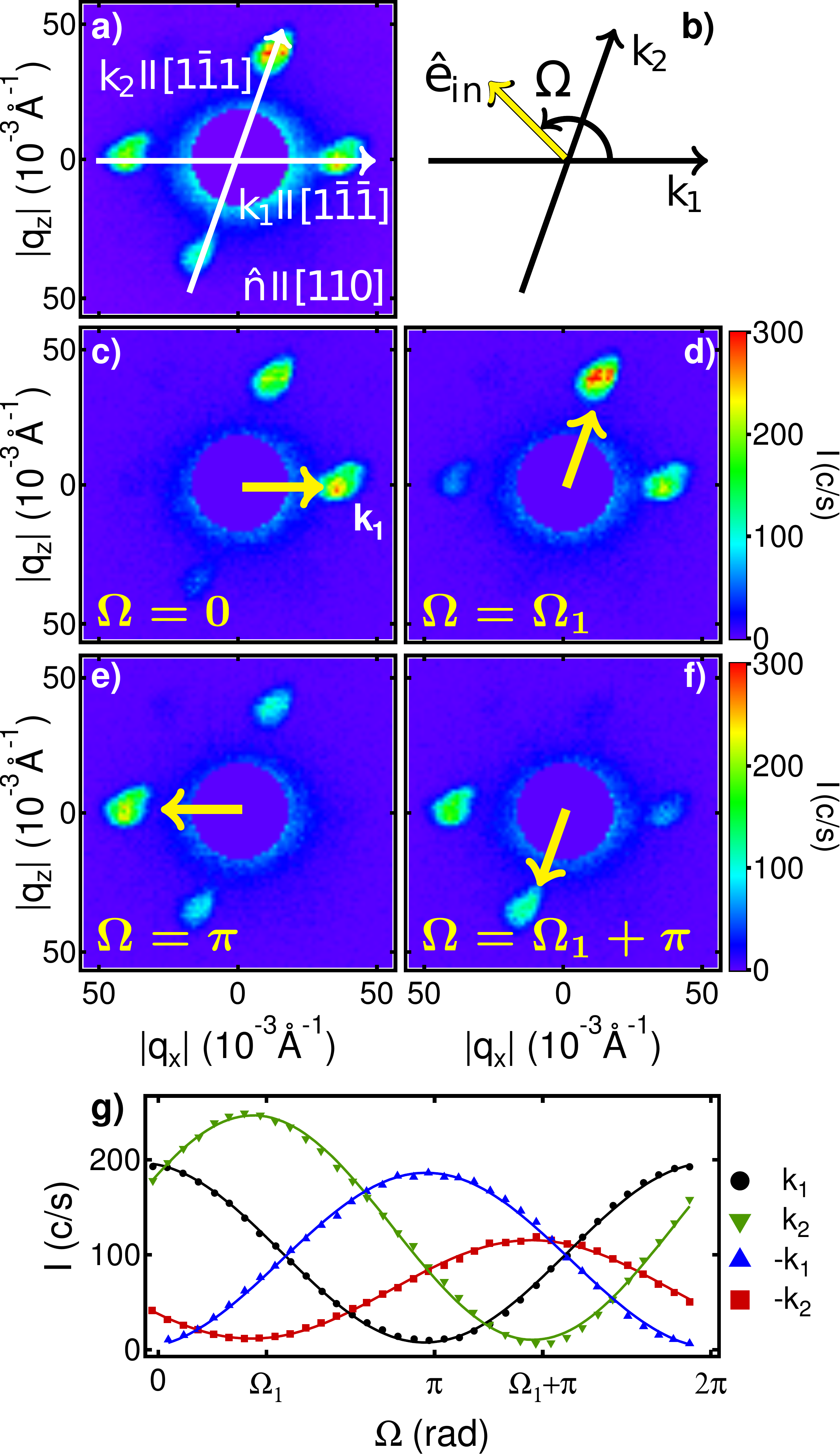}
\caption{
Operating principle of MiniMuPAD demonstrating the spin-selective Bragg scattering off helimagnetic order in MnSi. (a) Wave vectors, $\mathbf{k}_1$ and $\mathbf{k}_2$, of two different helimagnetic domains are within the plane orthogonal to the neutron beam $\hat n$. (b) The angle $\Omega$ defines the relative orientation of the spin polarisation of the incoming neutrons $\hat e_{\rm in}$ orthogonal to $\hat n$. (c) - (f) SANS patterns obtained for specific polarizations $\hat e_{\rm in}$. (g) Intensity of magnetic Bragg peaks at $\pm \mathbf{k}_{1,2}$ as a function of $\hat e_{\rm in}$, i.e., $\Omega$, where $\Omega_1 = \arccos\frac{1}{3}$. }
\label{figure2}
\end{figure}

For our study we used the MnSi single crystal investigated in Ref.\,\cite{janoschek2013fluctuation}. The sample was oriented by x-ray Laue diffraction such that two $\langle111\rangle$ axes, in the following denoted by $\boldsymbol{k}_1\parallel [1\bar{1}\bar{1}]$ and $\boldsymbol{k}_2\parallel [1\bar{1}1]$, were orthogonal to the neutron beam $\hat n\parallel [110]$, see Fig.~\ref{figure2}(a). The combination of the large window of the PCs with a PSD detector and the possibility for rotations of the sample with respect to its vertical $[1 \bar{1} 2]$ axis by an angle $\Phi=0^\circ,3^\circ,25^\circ$, see also Fig.~\ref{figure1}(c), allowed to track various points in reciprocal space going well beyond previous work.

For a physically transparent theoretical account of the spin-flip scattering in chiral magnets it is helpful to begin with the spin-resolved energy-integrated neutron scattering cross section, $\sigma_{\hat e_{\rm out}, \hat e_{\rm in}}(\mathbf{Q}) \equiv \frac{d \sigma}{d \Omega}$ with momentum transfer $\mathbf{Q}$, which describes the change of the spin eigenstate of the neutron from $|\hat e_{\rm in} \rangle$ to $|\hat e_{\rm out} \rangle$ with $\vec \sigma | \hat e_\alpha\rangle = \hat e_\alpha| \hat e_\alpha\rangle$. It comprises, in general, a nuclear and a magnetic contribution, as well as a nuclear--magnetic interference term \cite{Blume1963,Maleyev1962}. 
The spin-polarisation dependence of the cross-section, $\sigma(\mathbf{Q},\hat e_{\rm in}) = \sum_{\tau=\pm1} \sigma_{\tau \hat e_{\rm out}, \hat e_{\rm in}}(\mathbf{Q})$ may then be resolved, when scattering a polarised neutron beam. The magnetic contribution to $\sigma(\mathbf{Q},\hat e_{\rm in})$ consists thereby of a symmetric and an antisymmetric part $\sigma_{\rm mag}(\mathbf{Q},\hat e_{\rm in}) = \sigma^S_{\rm mag}(\mathbf{Q}) +  (\hat Q \hat e_{\rm in}) \sigma^A_{\rm mag}(\mathbf{Q})$, where the latter is weighted by the scalar product $\hat Q \hat e_{\rm in}$ with $\hat Q = \mathbf{Q}/Q$. The so-called {\it chiral fraction} $\eta$, is finally defined as $\eta(\mathbf{Q}) = \sigma^A_{\rm mag}(\mathbf{Q})/\sigma^S_{\rm mag}(\mathbf{Q})$, measuring the chirality of the magnetic scattering.

Finite values of $\sigma^A_{\rm mag}$ (and thus $\eta$) originate in the polarization-dependent scattering characteristics of chiral magnetic systems like MnSi. In the helimagnetically ordered phase of MnSi, the magnetisation may be described as $\vec M(\mathbf{r}) = M_{\rm hel} (\hat e_1 \cos \mathbf{k}\mathbf{r} + \hat e_2 \sin \mathbf{k}\mathbf{r})$, with wave vector $\mathbf{k}$ and orthonormal unit vectors $\hat e_i \hat e_j = \delta_{ij}$, with $i,j = 1,2$ orthogonal to $\mathbf{k}$; for a left-handed helix $\hat e_1 \times \hat e_2 = -\hat k$. The magnetic Bragg scattering by a magnetic helix possesses equal magnitudes of symmetric and antisymmetric contributions, $\sigma^S_{\rm mag}(\mathbf{k})|_{\rm Bragg} = \pm \sigma^A_{\rm mag}(\mathbf{k})|_{\rm Bragg}$ where the sign indicates the handedness of the magnetic helix with $\eta_{\rm Bragg}(\mathbf{k}) = 1$ being left- and $\eta_{\rm Bragg}(\mathbf{k}) = -1$ being right-handed. For left-handed helimagnetic order we thus expect $\sigma_{\rm mag}(\mathbf{k},\hat e_{\rm in})|_{\rm Bragg} \propto (1+(\hat k \hat e_{\rm in})) = 2 \cos^2(\delta \Omega/2)$, i.e., a sinusoidal dependence as a function of the angle $\delta\Omega$ enclosed by $\hat k$ and $\hat e_{\rm in}$.

In order to confirm the proper functioning of our SNP device this dependence was experimentally verified by rotating the polarisation $\hat e_{\rm in}$ of the incident neutron in the plane perpendicular to the beam as depicted in Fig.\,\ref{figure2}(b), where the orientation is denoted by the angle $\Omega = \angle(\hat e_{\rm in}, \mathbf{k}_1)$. Well below $T_{\rm c}$ weak cubic anisotropies align the helimagnetic wave vector $\mathbf{k}$ along crystallographic $\langle 111\rangle$ directions. In turn the intensity displays extrema whenever the polarisation $\hat e_{\rm in}$ aligns with a wave vector of a helimagnetic domain $\pm \mathbf{k}_{1,2}$ for $\Omega_0=0$, $\Omega_1=\arccos\frac{1}{3}$, $\Omega_2=\pi$ and $\Omega_3=\Omega_1 + \pi$. The associated intensity patterns at the PSD are shown in Fig.\,\ref{figure2}(c) through (f). Apart from a constant background, the variation of the intensity as a function of $\delta \Omega = \Omega - \Omega_n$, with $n=0,1,2,3$, shown in Fig.\,\ref{figure2}(g), may be fitted in perfect agreement with chiral Bragg scattering, underscoring the precision of the alignment of $\hat e_{\rm in}$. This suggests, in particular, a negligible interference between magnetic and nuclear contributions. Differences of the maximum peak intensity are thereby due to a small misalignment of the $\mathbf{k}_2$ crystallographic axis with respect to the SNP set-up. 

\begin{table}
\caption{Polarization matrix in the helimagnetically ordered phase of MnSi at 6.5\,K  
determined experimentally with our miniaturised SNP device.
\label{table1}}
\begin{tabular}{crrr}
\hline\noalign{\smallskip}
\hline\noalign{\smallskip}
$\mathbb{P}_{\alpha,\beta}$ & \multicolumn{1}{c}{$\beta=x$} &\multicolumn{1}{c}{$\beta=y$} &\multicolumn{1}{c}{$\beta=z$}\\
 \hline\noalign{\smallskip}
$\alpha=x$ & $-0.9660\pm0.0004$ & $-0.9692\pm0.0006$ & $-0.9664\pm0.0006$\\
$\alpha=y$ & $0.0115\pm0.0013$  & $0.0363\pm0.0018$  &  $0.0321\pm0.0018$\\
$\alpha=z$ & $ 0.0045\pm0.0013$ & $-0.079\pm0.0018$ &  $0.0182\pm0.0018$\\
\noalign{\smallskip}\hline
\noalign{\smallskip}\hline
\end{tabular}
\end{table}

Using in addition an analyser the fully spin-resolved scattering in the helical state may be discussed, where the polarisation matrix is defined as
\begin{align}
\mathbb{P}_{\alpha,\beta}(\mathbf{Q}) = 
\frac{\sigma_{\hat e_{\alpha},\hat e_{\beta}}(\mathbf{Q}) - \sigma_{-\hat e_{\alpha},\hat e_{\beta}}(\mathbf{Q}) }{\sigma_{\hat e_{\alpha},\hat e_{\beta}}(\mathbf{Q}) + \sigma_{-\hat e_{\alpha},\hat e_{\beta}}(\mathbf{Q}) }
\end{align}
and $\hat e_{\alpha}$ with $\alpha = x,y,z$ forms a right-handed triad. Far below $T_{\rm c}$  Bragg scattering prevails close to the magnetic wave vector $\mathbf{k}$ thus allowing to approximate $\sigma_{\hat e_{\rm out},\hat e_{\rm in}}(\mathbf{k}) \approx \sigma^{\rm mag}_{\hat e_{\rm out},\hat e_{\rm in}}(\mathbf{k})|_{\rm Bragg}$. In this case the polarisation matrix simplifies to $\mathbb{P}_{x,\beta}(\mathbf{k}) \approx -1$ for all $\beta = x,y,z$ and zero otherwise, where we assumed $\hat k = \hat e_x$ without loss of generality. The experimental values for $\mathbb{P}_{\alpha,\beta}(\mathbf{k}_1)$ recorded at 6.5\,K  accounting for the analyser efficiency are summarised in table\,\ref{table1}. The deviation by $\sim4\%$ from purely magnetic Bragg scattering may be attributed to the flip efficiency of the precession coils and  non-magnetic scattering by the coating of the precession coils and aluminium in the beam.

This brings us finally to the spin-resolved scattering close and above the helimagnetic transition at $T_c \approx 29$~K. Here the spin-polarisation of the in- and out-going neutron beam were longitudinal to the transferred momentum, $\sigma^\parallel_{\tau_{\rm out}, \tau_{\rm in}}(\mathbf{Q}) = \sigma_{\tau_{\rm out} \hat Q,\tau_{\rm in}\hat Q}(\mathbf{Q})$ with $\tau_{\rm out}, \tau_{\rm in} \in \{+1,-1\} \equiv \{\uparrow,\downarrow\}$. The spin-flip scattering $\sigma^\parallel_{\pm,\mp}(\mathbf{Q}) = \sigma^S_{\rm inc}(\mathbf{Q}) + \sigma^{S}_{\rm mag}(\mathbf{Q}) \mp \sigma^{A}_{\rm mag}(\mathbf{Q})$ is then goverened by critical chiral magnetism apart from an incoherent spin-flip background contribution $\sigma^S_{\rm inc}$. From the theory of chiral magnets \cite{Grigoriev2005} one expects the spin-flip scattering to assume the following simple form for $T>T_c$ 
\begin{align} \label{SpinFlip}
\sigma^{S}_{\rm mag}(\mathbf{Q}) \mp \sigma^{A}_{\rm mag}(\mathbf{Q}) =  \frac{\mathcal{A}\, k_B T}{(|\mathbf{Q}| \pm k)^2 + \kappa^2(T)},
\end{align}
where $k_B$ is the Boltzmann constant and $\mathcal{A}$ is a constant that depends on the form factor of MnSi. This expression applies to chiral systems tending to develop a left-handed helix with a pitch vector, $k>0$. The inverse correlation length, $\kappa(T)$, represents the point of contact with the different theoretical proposals of the helimagnetic transition that motivated our study. In particular, for very weak cubic anisotropies in a Brazovskii scenario chiral paramagnons develop isotropically and become soft on a sphere in momentum space as $\kappa(T) \to 0$, see Eq.~\eqref{SpinFlip} \cite{janoschek2013fluctuation}. These chiral paramagnons effectively display an one-dimensional character resulting in strong renormalizations of $\kappa(T)$ according to the cubic equation 
$\kappa^2/\kappa_{\rm Gi}^2 = \frac{T-T_{\rm MF}}{\tilde T_0} + \kappa_{\rm Gi}/\kappa$.
For MnSi $T_{\rm MF} \approx 30.5$ K is the expected mean-field transition temperature, where $\tilde T_0 \approx 0.5$ K and $\kappa_{\rm Gi} \approx 0.018$ \AA$^{-1}$ is the  inverse Ginzburg length \cite{janoschek2013fluctuation}. In turn, these strongly interacting paramagnons suppress the transition by $\Delta T = T_{\rm MF} - T_c \approx 1.5$~K driving it first order, with characteristic signatures in physical quantities around $T_{\rm MF}$.

\begin{figure}
\includegraphics[width=0.5\textwidth]{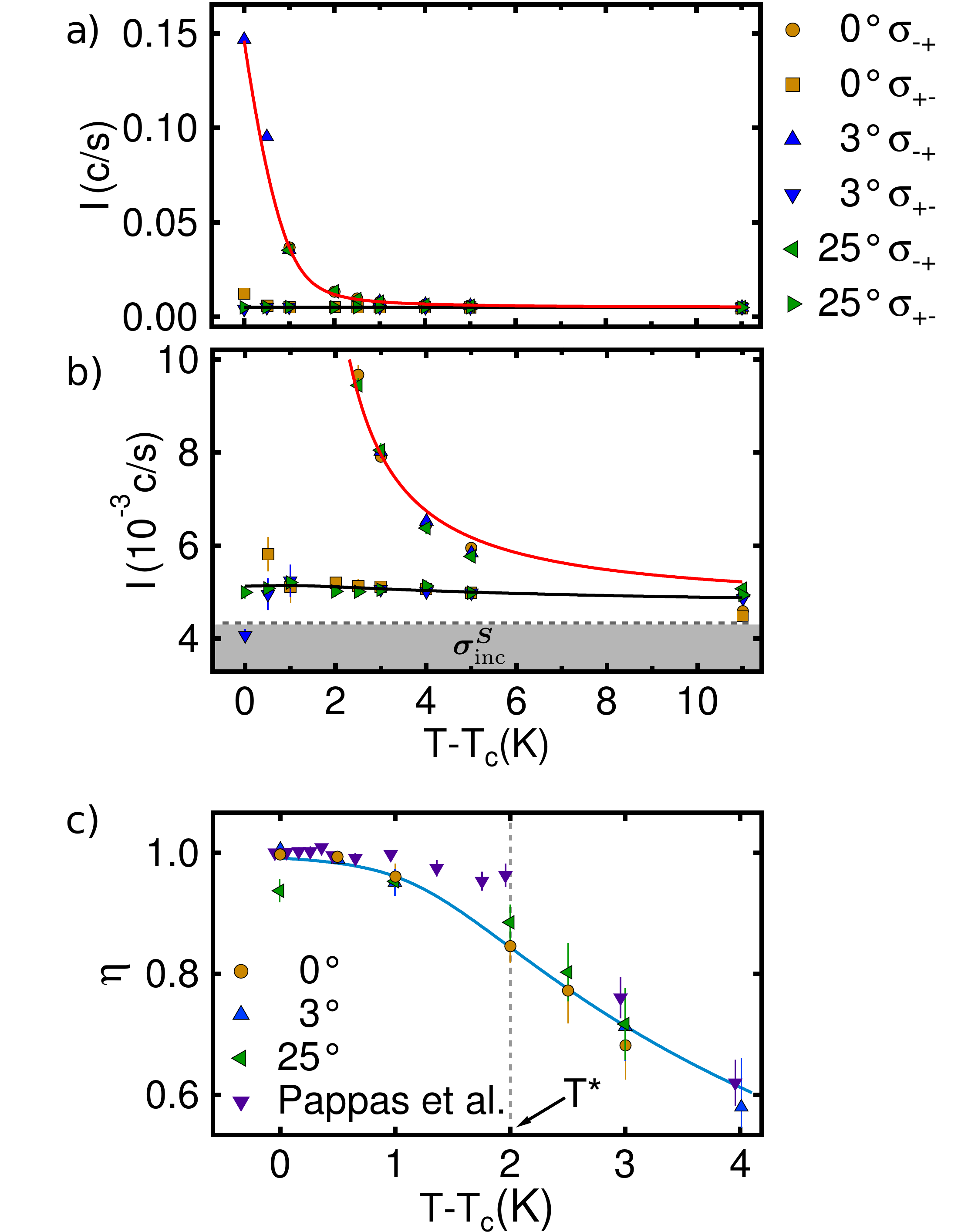}
\caption{(a) Temperature dependence of the spin-flip scattering cross section of MnSi $\sigma^\parallel_{\pm,\mp}(\mathbf{Q})$ with $|\mathbf{Q}| = k$ close to the critical temperature, $T>T_c$, for different orientations specified by the angle $\phi = 0^\circ, 3^\circ$, and $25^\circ$. For $\phi=0$, $\hat Q \parallel [1 \bar{1} \bar{1}]$ and other orientation are obtained by an anticlockwise rotation by $\phi$ around the $[1 \bar{1} 2]$ axis, see also Fig.~\ref{figure1}(c). Data for $\phi = 25^\circ$ was corrected for transmission. Panel (b) shows the same data but on a different intensity scale; the dashed line is the fitted incoherent spin-flip background value, $\sigma^S_{\rm inc}$. Panel (c) shows the chiral fraction $\eta$ of Eq.~\eqref{ChiralFraction}. The solid lines are fits to Brazovskii theory, see text, predicting a turning point of $\eta(T)$ at $T^*$. 
}
\label{figure3}
\end{figure}

Fig.~\ref{figure3}(a) and (b) display the temperature dependence of the spin-flip scattering 
$\sigma^\parallel_{\pm\mp}(\mathbf{Q})$ measured on a sphere with radius $|\mathbf{Q}| = k$ for different orientations $\hat Q = \mathbf{Q}/Q$. Approaching $T_c$, the chiral magnetic ordering indeed develops isotropically resulting in a negligible dependence on $\hat Q$. Here $\sigma^\parallel_{-+}$ reflects the strong $T$-dependence of $\kappa$ close to $T_c$, while $\sigma^\parallel_{+-}$ is barely temperature dependent as it is suppressed by the additional factor $4k^2$ in the denominator of Eq.~\eqref{SpinFlip}. Using the published results for $\kappa(T)$ and $k\approx 0.039$ \AA$^{-1}$ obtained in the same sample \cite{janoschek2013fluctuation}, we are left with a single fitting parameter, namely the magnitude $\mathcal{A}$ in Eq.~\eqref{SpinFlip}, in addition to a temperature and $\mathbf{Q}$-independent incoherent background $\sigma^S_{\rm inc}$ shown as the dotted line in Fig.~\ref{figure3}(b). We find a remarkably good fit for both cross sections as shown by the solid lines.

Subtracting the incoherent background $\sigma^S_{\rm inc}$ determined experimentally, we obtain the chiral fraction 
\begin{align} \label{ChiralFraction}
\eta \equiv \frac{\sigma^{A}_{\rm mag}(\mathbf{Q})}{\sigma^{S}_{\rm mag}(\mathbf{Q})}\Big|_{|\mathbf{Q}|=k} = \frac{1}{1 + \kappa^2(T)/(2k^2)}
\end{align}
shown in Fig.~\ref{figure3}\,(c). It is essential to note that the experimental values depend sensitively on $\sigma^S_{\rm inc}$ (likewise the error bars of $\eta$ derive mainly from $\sigma^S_{\rm inc}$). Within the error bars, however, the chiral fraction is in very good agreement with the Brazovskii theory of $\kappa(T)$.  In particular, $\eta(T)$  displays a characteristic point of inflection at a temperature $T^*-T_c \approx 2$ K.
It is finally instructive to note that $\eta(T)$ reported by Pappas {\it et al.} \cite{Pappas2009,Pappas2011}
differs substantially from up to $\sim2\,{\rm K}$ above $T_{\rm c}$ as shown in Fig.~\ref{figure3}(c). Based on the information given in Refs.\,\cite{Pappas2009,Pappas2011} we strongly suspect that this difference is due to an overestimate of $\sigma^S_{\rm inc}$. 

In conclusion, we have investigated the critical spin-flip scattering with an emphasis on the chiral fraction $\eta$ close to the helimagnetic transition in MnSi. For our study we have developed a miniaturised, low-cost SNP device for very fast experiments at scattering angles up to $15^{\circ}$. Considering carefully the importance of incoherent background scattering we find excellent quantitative agreement of the temperature dependence of the chiral fraction $\eta$ at various sample orientations with the Brazovski scenario of a fluctuation-induced first order transition. Our study thereby provides for the first time a quantitative connection of $\eta$ with elastic neutron scattering as well as the magnetisation, susceptibility and specific heat \cite{janoschek2013fluctuation}, completing a remarkably comprehensive account in
a minimal model that does not require any additional phenomenological parameters.

We wish to thank A. Rosch, K. Pappas and S. Grigoriev for helpful discussions. JK acknowledges financial support through the TUM Graduate School. Financial support through DFG TRR80 and ERC-AdG (291079 TOPFIT) are gratefully acknowledged.

\bibliography{BibDatenbank}
\end{document}